\documentclass[aps,preprint,showpacs,preprintnumbers,amsmath,amssymb]{revtex4}
\usepackage[dvips]{graphicx}
\usepackage{float}
\begin{document}

\title{Production of charged scalar particles with well defined angular momentum in de Sitter spacetime}
\author{Mihaela-Andreea B\u aloi}

\email{mihaela.baloi@upt.ro}
 \affiliation{Politehnica University of Timi\c soara,  \\V. P\^ arvan
 Avenue 2, RO-300223 Timi\c soara,  Romania}

\begin{abstract}
We study the production of charged scalar particles with well defined angular momentum, in the presence of an external Coulomb field on de Sitter expanding universe. This process of particle production is studied as a perturbative phenomenon applying the theory of quantum electrodynamics on de Sitter spacetime developed in \cite{1}-\cite{3}. The probability of particle production is computed and our results show the dependence of the probability in terms of the orbital quantum number. Our graphical results show that, the most probable transitions are those that generate particles with small angular momentum. Is obtained also that the momentum is not conserved and that the probability of particle production is significative for large values of the Hubble's constant, corresponding to the early universe.
\end{abstract}

\maketitle

\section{Introduction}
In the last years important progresses have been made regarding the construction of a quantum field theory on curved spacetimes. Solutions of the free field equations were found and a theory of interacting fields with plane waves solutions was proposed.  Spherical wave solutions of field equations on curved spacetimes were found in \cite{26}-\cite{4}. In \cite{26} was solved the Dirac equation in spherical coordinates, coupled with an external potential, based on the separation of variables method. In \cite{27}, spherical spinor solutions of the free Dirac equation were obtained on de Sitter expanding universe in the chart $\{t,r,\theta,\phi\}$. In this paper the authors solve the Dirac equation in local frames and show that these solutions are eigenfunctions of the total angular momentum operator. Another spherically symmetric solution of the Dirac field in the chart $\{t,r,\theta,\phi\}$, was obtained by Shishkin using the method of the separation of variables \cite{28}. The problem of finding solutions of the Klein-Gordon equation in different charts on a curved background was treated in \cite{4}-\cite{6}. In \cite{4} was obtained the exact solution in spherical coordinates and is constructed the quantum theory of the free scalar field on de Sitter expanding universe.\\
The maximally symmetric solution of the Einstein's equations, namely the de Sitter spacetime, is one of the most exploited subjects in the literature when comes to construct interactions between quantum matter in the presence of gravitation. This is because the de Sitter geometry is suitable for computations needed in the field theory (ten Killing vectors) and also because the astronomical observations show that our universe has an accelerated expansion. Eight decades ago, E. Schr\"{o}dinger publishes a paper where he studies the Klein-Gordon equation on curved spacetimes and reaches the conclusion that, the expansion of the space could produce quantum matter \cite{7}. Since Schr\"{o}dingr's  work this topic received an increasingly attention in the literature.\\
A particular topic that has attracted the attention of  many authors is the production of particles in the presence of external electromagnetic fields on curved spacetimes and this subject is studied as well as a perturbative and as a nonperturbative phenomenon \cite{1}\cite{14}-\cite{13}. In the case of the electric fields were obtained important results such as, the particles are most probable emitted parallel to the direction of the electric field \cite{8}\cite{09}. Recent studies also approached  the problem of fermions production and the production of the spinless particles in the presence of a magnetic field, which is produced by a magnetic dipole in spatially flat Friedmann-Lema\^itre-Robertson-Walker backgrounds \cite{14}-\cite{19}.  In this case it is showed that the created particles are most probably emitted perpendicularly to the direction of the magnetic field. This subject is interesting because the astronomical observations show that the galaxies have a proper magnetic field which is believed to own its existence in the early universe. But it is important to mention that, all these cases discuss the creation of particles described by the plane waves solutions of the field equations in de Sitter geometry. Therefore in all of these cases we have information regarding the conservation of charge, the conservation of momentum  and the conservation of helicity, if we discuss about the production of Dirac particles.\\
The conservation of charge is the natural conservation law of nature and cannot be violated in QED processes. Thus the immediate consequence is that the particles must be produced in pairs. Regarding the conservation of momentum and helicity, these two physical quantities can or cannot be conserved, depending on the conditions in which we are studying the creation of particles \cite{3}\cite{14}.\\
It is then an interesting issue to study what happens if the produced particles also have a well determined angular momentum. Then all the quantities such as the transition amplitudes and the probabilities will depend on the angular momentum. As far as we known, this subject is poorly treated in the literature mostly because of the mathematical complexity of such a topic. The aim of this paper is to study the production of charged scalar particles described by spherical wave solutions of the Klein-Gordon equation in the presence of a Coulomb field, on de Sitter spacetime. More precisely, the intention of this paper is to obtain the probability dependence in terms  of the angular momentum and to extract the physical consequences of these results. \\
The paper is organized as follows: in the second section we discuss about the fundamental solutions of the free field equations, in the third section we give the final result of the probability of pair production and we discuss about the physical consequences of our results. Then in section four we graphically analyze the probability of particle production and in section five we discuss about the limit cases of our results. In section six we present our conclusions and future perspectives. In the Appendix we give the basic steps that are needed in order to solve the integrals at which one arises in this paper.

\section{Spherical wave solutions}
The line element of the de Sitter expanding geometry in the moving chart with spherical coordinates $\{t, r, \theta, \varphi\}$ reads:
\begin{equation}
ds^2= dt^2-e^{2\omega t}(dr^2 + r^2d\theta^2+ r^2\sin\theta^2+ d\varphi^2),
\end{equation}
 where $\omega> 0$ is the Hubble's constant.
The theory of free scalar field with plane wave solutions on de Sitter spacetime is constructed in Refs.\cite{5}\cite{6}. In this paper we are interested to study the production of charge scalar particles with well defined angular momentum. Thus it is important to mention that the solution of the Klein-Gordon equation in the chart with spherical coordinates has the same time modulation as the plane wave solutions obtained in the Cartesian chart $\{t,\vec{x}\}$ \cite{6}, given by the Hankel functions. The radial part depends on Bessel J functions and the angular part contains the well-known spherical harmonics $Y_{l,m}$ from flat space theory. Therefore
the positive frequency solution of the Klein-Gordon equation in the angular momentum basis reads \cite{4}:
\begin{equation}\label{sol2}
f_{p,l,m}(t, r, \theta, \varphi)=\frac{1}{2\sqrt{r}}\sqrt{\frac{p\pi}{\omega}}e^{-3\omega t/2}e^{-\pi\mu/2}H_{i\mu}^{(1)}\left(\frac{p}{\omega}e^{-\omega t}\right)J_{l+\frac{1}{2}}(pr)Y_{l,m}(\theta, \varphi),
\end{equation}
where $\mu= \sqrt{\left(\frac{m}{\omega}\right)^2-\frac{9}{4}}$ and $\frac{m}{\omega}> \frac{3}{2}$. In expression $(\ref{sol2})$ the modulus of the momentum vector is denoted by $|\vec{p}|= p$,  the orbital quantum number which quantize the angular momentum is denoted by $l= 0,1,2,...$, and $m={-l,...\,,+l}$ $((2l+1)$ values) is the magnetic quantum number which quantize the projection of the angular momentum on axis. The negative frequency solutions $f^{*}_{p,l,m}(t, r, \theta, \varphi)$ are obtained applying the complex conjugation operation on $f_{p,l,m}(t, r, \theta, \varphi)$ . The set of the fundamental solutions
$\{f_{p,l,m}(t, r, \theta, \varphi),\, f^{*}_{p,l,m}(t, r, \theta, \varphi)\}$ which give the quantum modes of the massive spinless particles are correctly normalized accordingly to the scalar product of the Klein-Gordon theory:
\begin{equation}
i\,e^{3\omega t}\int_{0}^{\infty}dr r^2\int d\Omega f^{*}_{p,l,m}(t, r, \theta, \varphi)\stackrel{\leftrightarrow}{\partial_{t}}f_{p',l',m'}(t, r, \theta, \varphi)=\delta(p-p')\,\delta_{l, l'}\,\delta_{m, m'}.
\end{equation}
The external field used in our analysis is a Coulomb field. In order to establish the expression of the Coulomb potential on de Sitter spacetime we take into account that the Maxwell's equations are invariant under the conformal transformations and also that the de Sitter metric is conformal with the Minkowski one \cite{2}. Therefore, in the natural frame, the Coulomb potential is given by the following expression \cite{2}:
\begin{equation}
A^{0}(r)= \frac{Ze}{r} e^{-2\omega t}.
\end{equation}
At this section we introduced the basic elements in order to compute the amplitude and the probability of transition corresponding to the process $vacuum\rightarrow \varphi+ \varphi^{*}$, in the presence of the Coulomb field. The case $\varphi+ \varphi^{*}\rightarrow vacuum$ represents the situation in which the pair annihilates.

\section{Probability of pair production}
The expression of the transition amplitude in the first order of perturbation theory on de Sitter spacetime is \cite{1}:
\begin{equation}
\mathcal{A}_{i\rightarrow f}= -e\int \sqrt{-g}\left[f_{\vec{p}}^{*}(t,\vec{x})\stackrel{\leftrightarrow}{\partial_{\mu}}f_{\vec{p}\,'}^{*}(t,\vec{x})\right]A^{\mu}(x)\,d^4x,
\end{equation}
where $\sqrt{-g}= e^{3\omega t}$. It is important to mention that, the free field theory and the theory of interactions is constructed on de Sitter expanding universe considering that, both the scalar and electromagnetic fields are minimally coupled with de Sitter background.
The expression of the transition amplitude, taking into account that $d^4x= r^2 dr\,d\Omega\,dt$ becomes:
\begin{equation}\label{a1}
\mathcal{A}_{i\rightarrow f}= -e\int r^{2}\sqrt{-g}\left[f_{p,l,m}^{*}(t, r, \theta, \varphi)\stackrel{\leftrightarrow}{\partial_{t}}f_{p',l',m'}^{*}(t, r, \theta, \varphi)\right]A^{0}(r) dr\, d\Omega\, dt,
\end{equation}
where  $d\Omega=\sin\theta \,d\theta\,d\varphi$, $\theta\in[0, \pi];\,\varphi\in[0, 2\pi]$. At this section, of a particular interest it will be to calculate expressions of the form $(\ref{a1})$, because in this way we can obtain the analytical formula for the processes that imply particle production and further we can graphically analyze such expressions. So let us first shortly discuss about the basics steps that are needed in order to compute expressions like $(\ref{a1})$. Firstly, after replacing the solutions of the spherically Klein-Gordon equation and the Coulomb potential in the expression $(\ref{a1})$ we obtain three type of integrals that are discussed in the Appendix A, namely the spatial integral (\ref{s1}), the angular integral (\ref{u1}) and respectively the temporal integral (\ref{t1}), which carries the influence of the de Sitter gravity upon this process through the expansion parameter $\omega$.
At an intermediary step in the calculation the expression of the amplitude is:
\begin{eqnarray}\label{a3}
\mathcal{A}_{i\rightarrow f}&=&-\frac{Ze^2\pi\omega\sqrt{p\,p'}}{4} e^{-\pi\mu}\int_{0}^{\infty}dz z^2\left[p\,\frac{\partial H_{-i\mu}^{(2)}(pz)}{\partial (pz)}H_{-i\mu}^{(2)}(p'z)-p'H_{-i\mu}^{(2)}(pz)\frac{\partial H_{-i\mu}^{(2)}(p'z)}{\partial (p'z)}\right]\nonumber\\
&&\times\int_{0}^{\infty}dr J_{l+\frac{1}{2}}(pr)J_{l'+\frac{1}{2}}(p'r)\int d\Omega \,Y_{l,m}^{*}(\theta,\varphi)Y_{l',m'}^{*}(\theta',\varphi'),
\end{eqnarray}
where the new variable of integration in temporal integral is $z=e^{-\omega t}/\omega$ and the bilateral derivation in amplitude is done.
It is observed that, expression $(\ref{a3})$ contains derivatives of Hankel functions with respect to the argument of these functions.
In order to solve integrals of the form given in the above expression of the amplitude we have to use a recurrence relation between Hankel functions and also we have to use general expressions of integrals that imply products of Hankel and Bessel J functions. But we will resume here only to give the final results of our analytical calculations and to discuss the physical consequences that are obtained and the detail steps that are needed to solve these kind of integrals are presented in the Appendix A.
Collecting all the results and notations we can finally establish the general form of the amplitude of the process $vacuum\rightarrow \varphi + \varphi^{*}$ in the presence of a Coulomb potential type, which is given by the following expression:
\begin{eqnarray}\label{a2}
\mathcal{A}_{i\rightarrow f}&=& -\frac{i(-1)^{m'}Ze^2\omega}{2\pi }\,\biggl\{\frac{1}{p^2}\theta(p-p')\left[f_{\mu}\left(\frac{p'}{p}\right)- g_{\mu}\left(\frac{p'}{p}\right)\right]h_{l'}\left(\frac{p'}{p}\right)\nonumber\\
&&+\frac{1}{p'^2}\theta(p'-p)\left[ f_{\mu}\left(\frac{p}{p'}\right)-g_{\mu}\left(\frac{p}{p'}\right)\right]h_{l}\left(\frac{p}{p'}\right)\biggl\}\delta_{l,l'}\delta_{m,-m'}.
\end{eqnarray}
One of the first results that are obtained from expression $(\ref{a2})$ is that the momentum is not a conserved quantity in this process. This is because the transition amplitude depends on the Heaviside step functions. Therefore we can have two situations in which particles are produced: either $p> p'$ or $p'> p$. It turns out that the momentum conservation law is broken in all cases of particle production in the presence of external fields, in de Sitter geometry. It was argued before that, the momentum conservation law is broken because of the presence of the external fields in de Sitter spacetime \cite{14}. Therefore when we study processes in the absence of external fields like for example: $vacuum\rightarrow\varphi+ \varphi^{*}+ \gamma$ (the production of the triplet scalar particle, scalar antiparticle and one photon) or $\varphi+ \varphi^{*}\rightarrow\gamma$ (annihilation of the scalar particles into a photon), the momentum is conserved (see Refs.\cite{1}\cite{29}). Another important result that is obtained from $(\ref{a2})$ is that the orbital quantum numbers of particle and antiparticle are equal, i.e. $l= l'$.  Also for having a nonzero transition amplitude it is necessary for the magnetic quantum number of the particle to take the same value as the magnetic quantum number of the antiparticle, but opposite as sign $(m= -m')$. This result can be interpreted as follows: because we study transitions from a vacuum state to a two-particle state and in order for the  projection on axis of angular momentum to be conserved, it is necessary that $m= -m'$. Since the angular momentum measures the rotation of the momentum vector on trajectory then our result with $m= -m'$, shows that the particle is spinning in opposite sense relatively to the antiparticle.

In general, in the case of scalar field, the probability of particle production is calculated as the square modulus of the transition amplitude (\ref{a1}):
\begin{equation}
\mathcal{P}_{i\rightarrow f}= \left|\mathcal{A}_{i\rightarrow f}\right|^2.
\end{equation}
Computing the square modulus of expression $(\ref{a2})$ we obtain the final result of the probability of scalar particle production with spherical waves:
\begin{eqnarray}\label{pr}
\mathcal{P}_{i\rightarrow f}&= & \frac{(-1)^{2m'}Z^{2}e^{4}\,m^2}{4\pi^2\,k^2}\biggl\{\left[\frac{1}{p^4}\theta(p-p')\left|f_{\mu}\left(\frac{p'}{p}\right)-g_{\mu}\left(\frac{p'}{p}\right)\right|^{2}h_{l'}^{2}\left(\frac{p'}{p}\right)\right.\nonumber\\
&&\left.+ \frac{1}{p'^{4}}\theta(p'-p)\left|f_{\mu}\left(\frac{p}{p'}\right)-g_{\mu}\left(\frac{p}{p'}\right)\right|^{2}h_{l}^{2}\left(\frac{p}{p'}\right)\right]
\biggl\}\delta_{l,l'}\delta_{m,-m'}.
\end{eqnarray}
The result $(\ref{pr})$ depends on the physical parameters ratio of the momenta of the produced particles $p'/p$, ratio mass of the particle per expansion factor $k= m/\omega$ and on the quantum numbers $l= l'$, respectively $m= -m'$. Because the probability dependence on the physical parameters is not so simple and is given by terms which contain special functions (Gauss hypergeometric function, Bessel J function) it is usefully to study such expressions using graphical methods. Therefore the purpose of the next section is to graphically analyze the probability in terms of the parameters $p'/p$, $k= m/\omega$ and $l$.

\section{Graphical results}
The topic of this section is to graphically study the probability in terms of parameter $k= m/\omega$, fixing the momenta ratios $p'/p$, for different values of the orbital quantum number $l=0, 1, 2, 3$. Another interesting issue that we want to approach at this section is to represent graphically the probability in terms of $l$ for fixed ratios $p\,'/p$ and fixed values of $k$.  As one can see from (\ref{pr}), the probability is dependent on the quantum number $l$. The probability equation could help us to establish a relation between the values of $l$ and the most probable transitions when the analysis is done in terms of small/large values of the Hubble's parameter $\omega$ and the momenta ratios of the produced particles is fixed. We mention that, the software used to plot our probabilities is Maple.
\begin{figure}[H]
\includegraphics[scale=0.35]{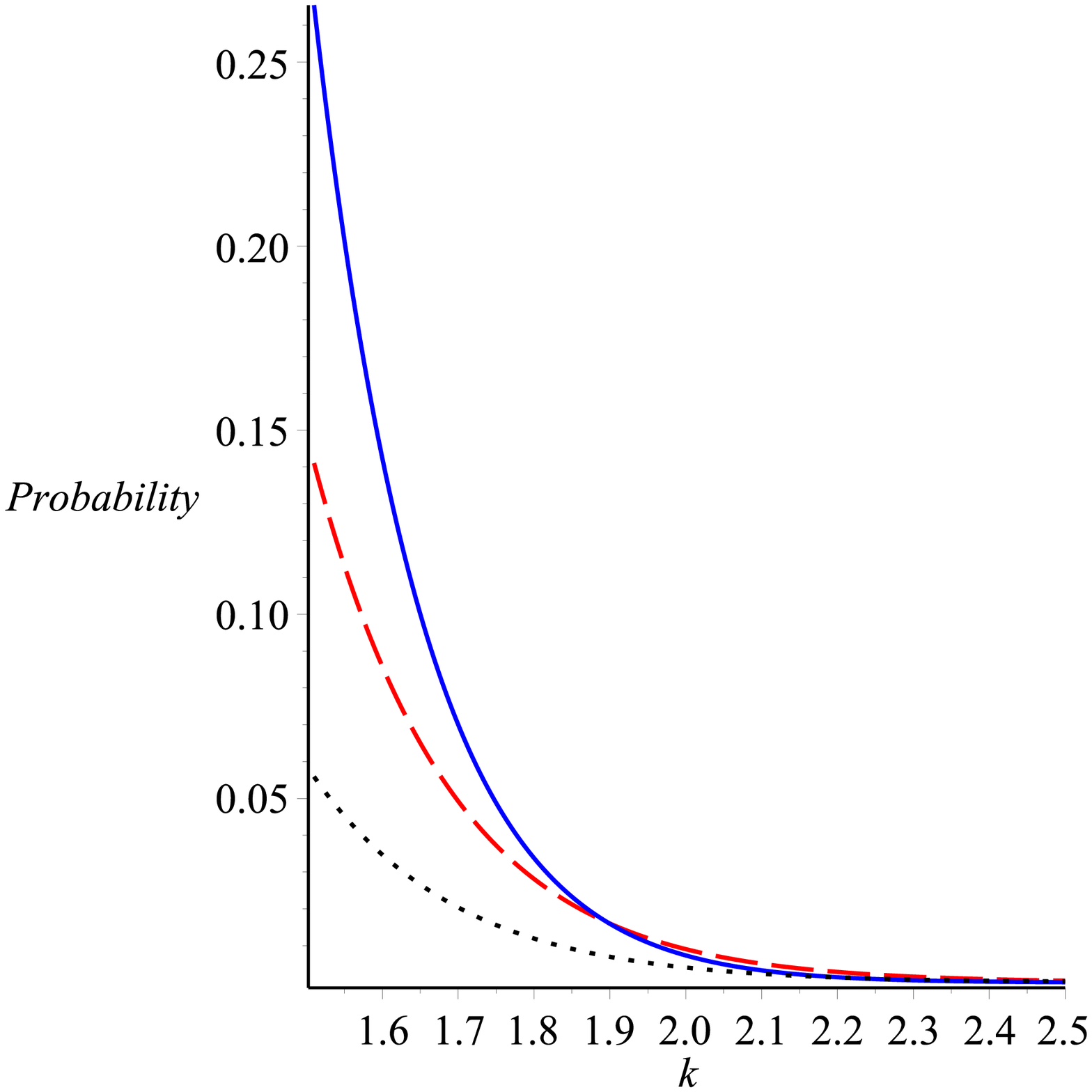}
\includegraphics[scale=0.35]{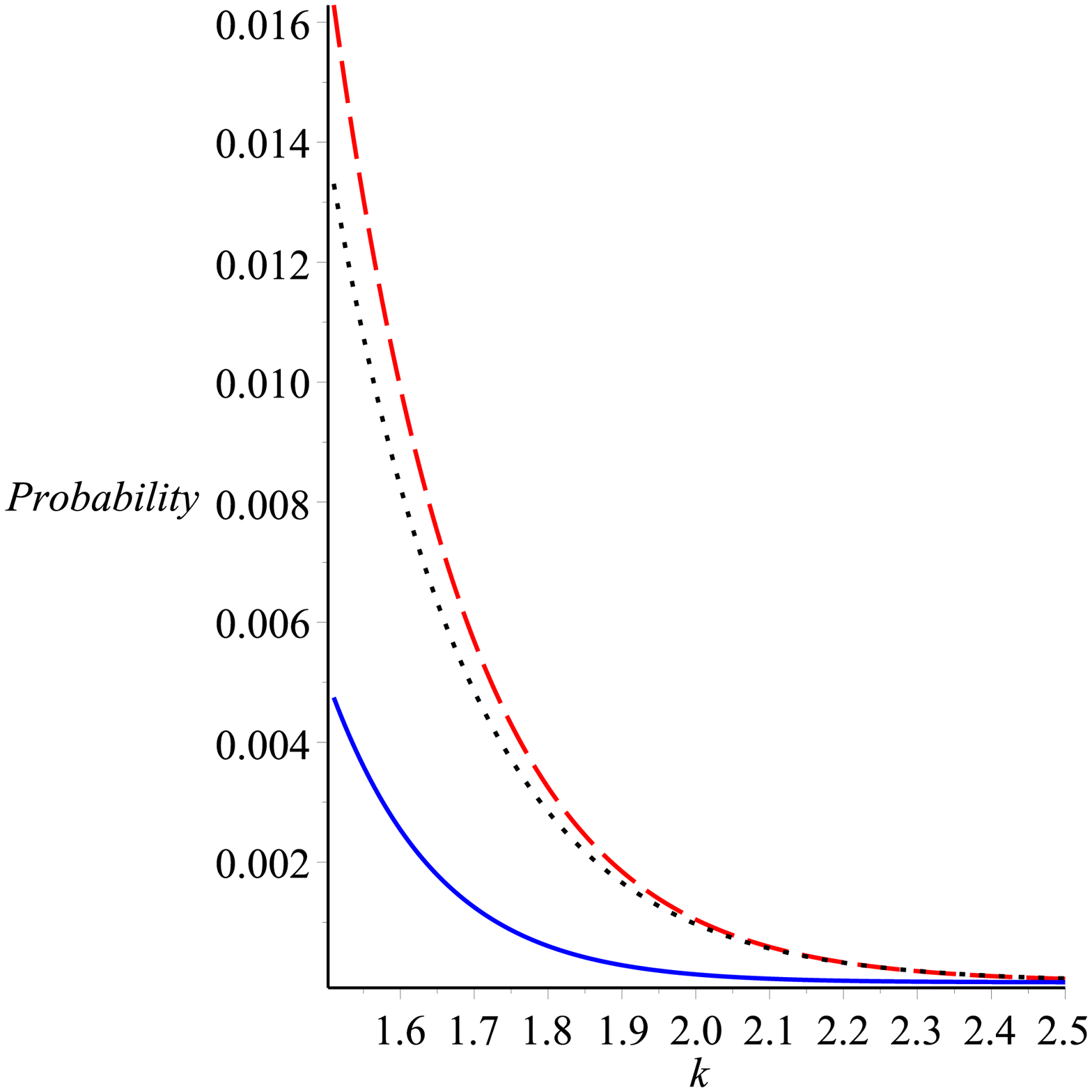}
\caption{$\mathcal{P}$ as a function of $k=\frac{m}{\omega}$. The numerical values of the momenta ratios are: $p\,'/p=0.2$ (solid line), $p\,'/p=0.5$ (dashed line), $p\,'/p=0.7$ (point line). In the left figure $l=0$ and in the right figure $l=1$.}
\label{fg1}
\end{figure}

\begin{figure}[H]
\includegraphics[scale=0.35]{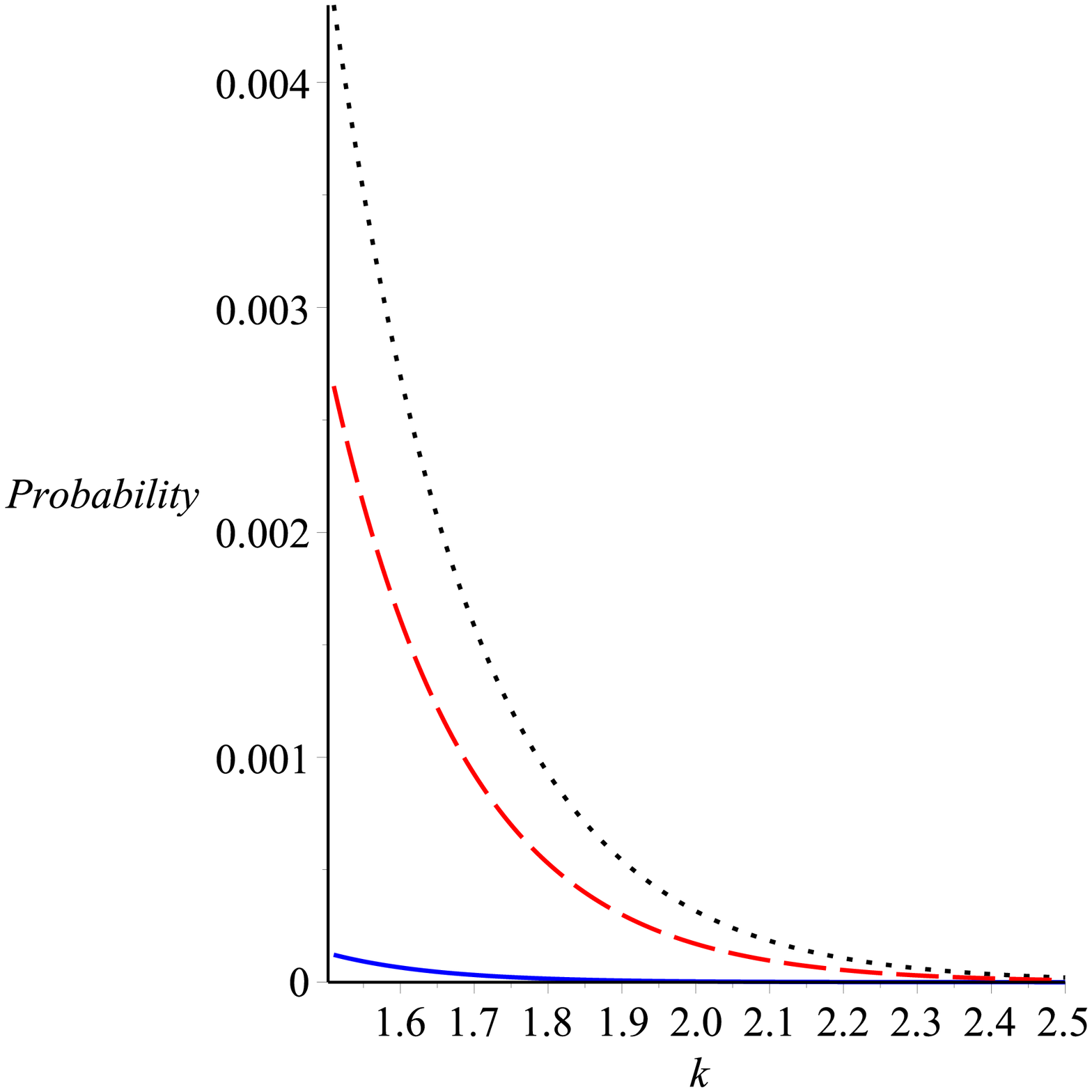}
\includegraphics[scale=0.35]{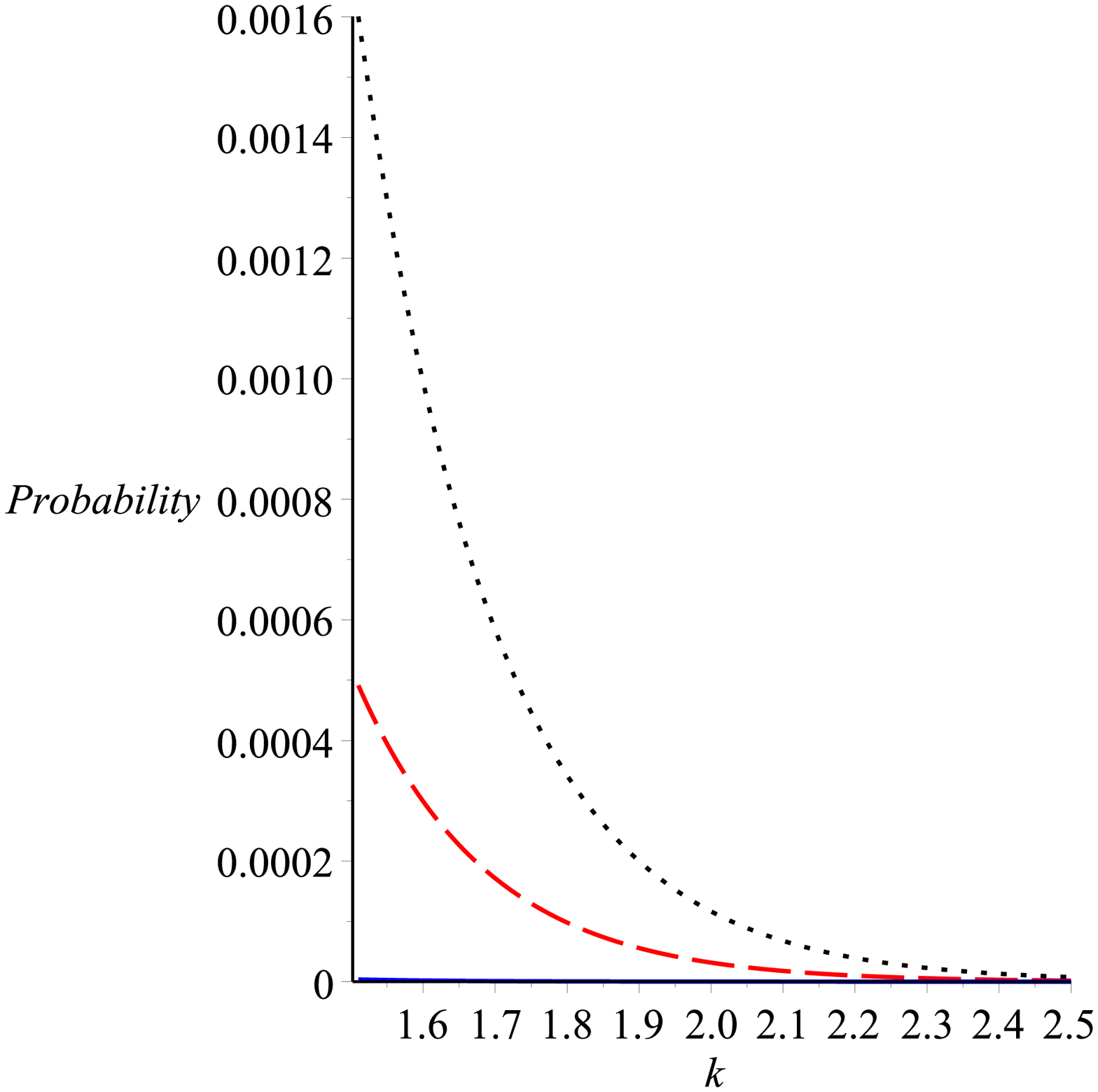}
\caption{$\mathcal{P}$ as a function of $k=\frac{m}{\omega}$. The numerical values of the momenta ratios are: $p\,'/p=0.2$ (solid line), $p\,'/p=0.5$ (dashed line), $p\,'/p=0.7$ (point line). In the left figure $l=2$ and in the right figure $l=3$.}
\label{fg2}
\end{figure}
The first observation from Figs.(\ref{fg1})-(\ref{fg2}) is that, the probability is large when the parameter $k$ is small (large expansion factor) and the orbital quantum number is zero $l=0$. Then if we increase the values of the orbital quantum number $l=1,\, 2, \,3$ the probabilities decrease, but still remain significative for small values of $k$. This behaviour is kept for all  values of the parameters $p,\, p' $and $l$. For the same values of momenta parameters and for increasing values of the orbital quantum number, the probability for the case in which the momenta ratio is close to one became more larger in comparison with the situation in which, for example $p'/p= 0.2$ (see Figs.(\ref{fg1})-(\ref{fg2})). But the variation of the probability in terms of the quantum number $l$ shows that, the probability rapidly decreases for large values of $l$. So, the most probable case of particle production is the case in which the angular momentum of those particles is small. The probabilities of pair production for the cases in which the orbital quantum number is large are significative if the momenta ratios are close to one.\\
Another observation is that, the functions which enter in the definition of the probability are very convergent and this behaviour is revealed by the fact that the probability of pair production decreases rapidly towards zero when the parameter $k\rightarrow \infty$.
\newline
In the following graphs we considered small values of the ratios of the momenta, but we kept for the orbital quantum number the same values as in the previous case $l=0, 1, 2, 3,$ because we want to see how the behaviour of the functions $f_{\mu}\left(\frac{p'}{p}\right)$, $g_{\mu}\left(\frac{p'}{p}\right)$ and $h_{l}\left(\frac{p'}{p}\right)$ that define the probability is changed. The results can be seen in the graphs below.
\begin{figure}[H]
\includegraphics[scale=0.35]{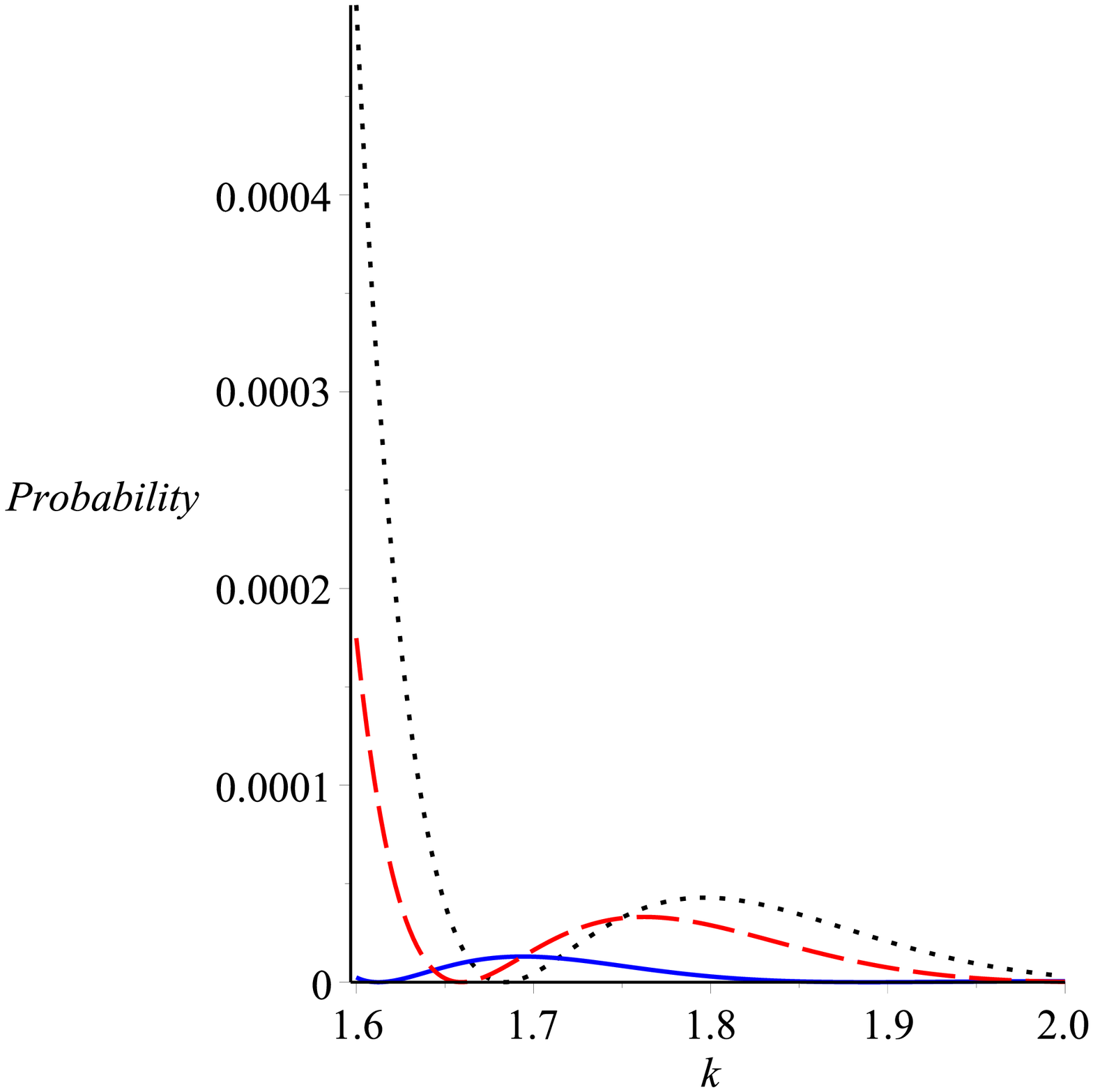}
\includegraphics[scale=0.35]{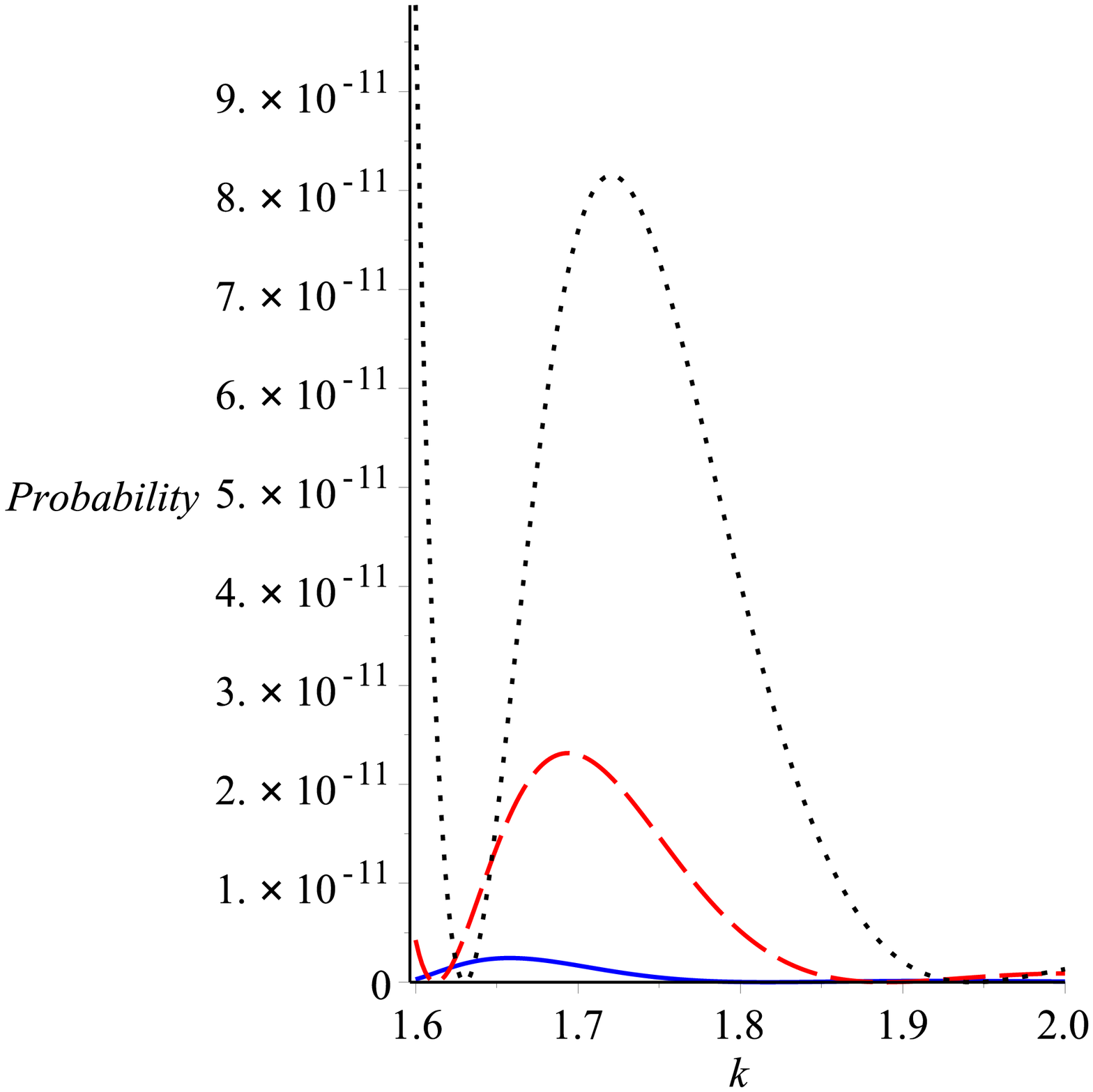}
\caption{$\mathcal{P}$ as a function of $k=\frac{m}{\omega}$. In the left figure the numerical values of the momenta ratios are: $p\,'/p=0.002$ (solid line), $p\,'/p=0.005$ (dashed line), $p\,'/p=0.007$ (point line) and $l=0$. In the right figure the numerical values of the momenta ratios are: $p\,'/p=0.001$ (solid line), $p\,'/p=0.002$ (dashed line), $p\,'/p=0.003$ (point line) and $l=1$.}
\label{fg3}
\end{figure}

\begin{figure}[H]
\includegraphics[scale=0.35]{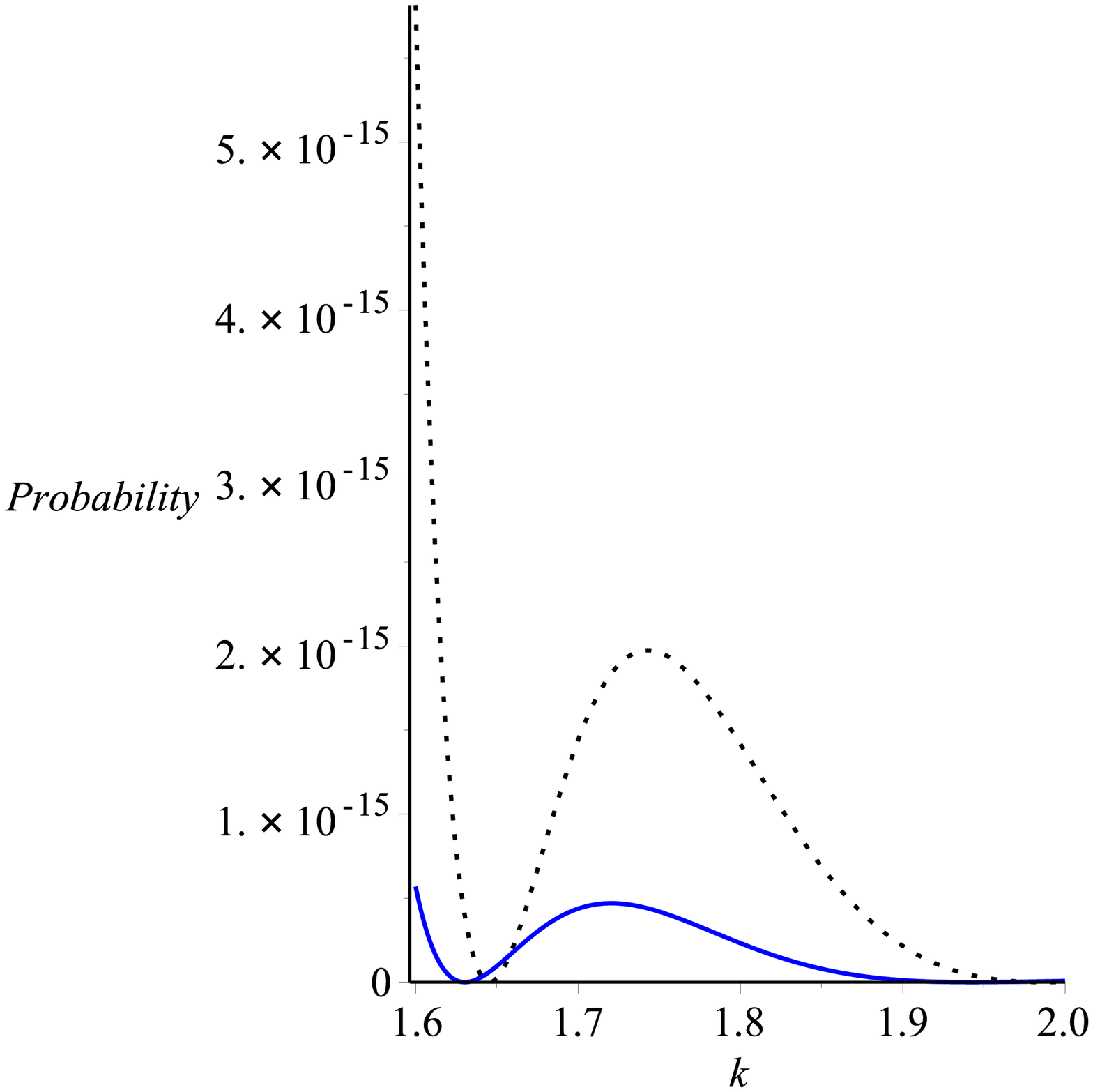}
\includegraphics[scale=0.35]{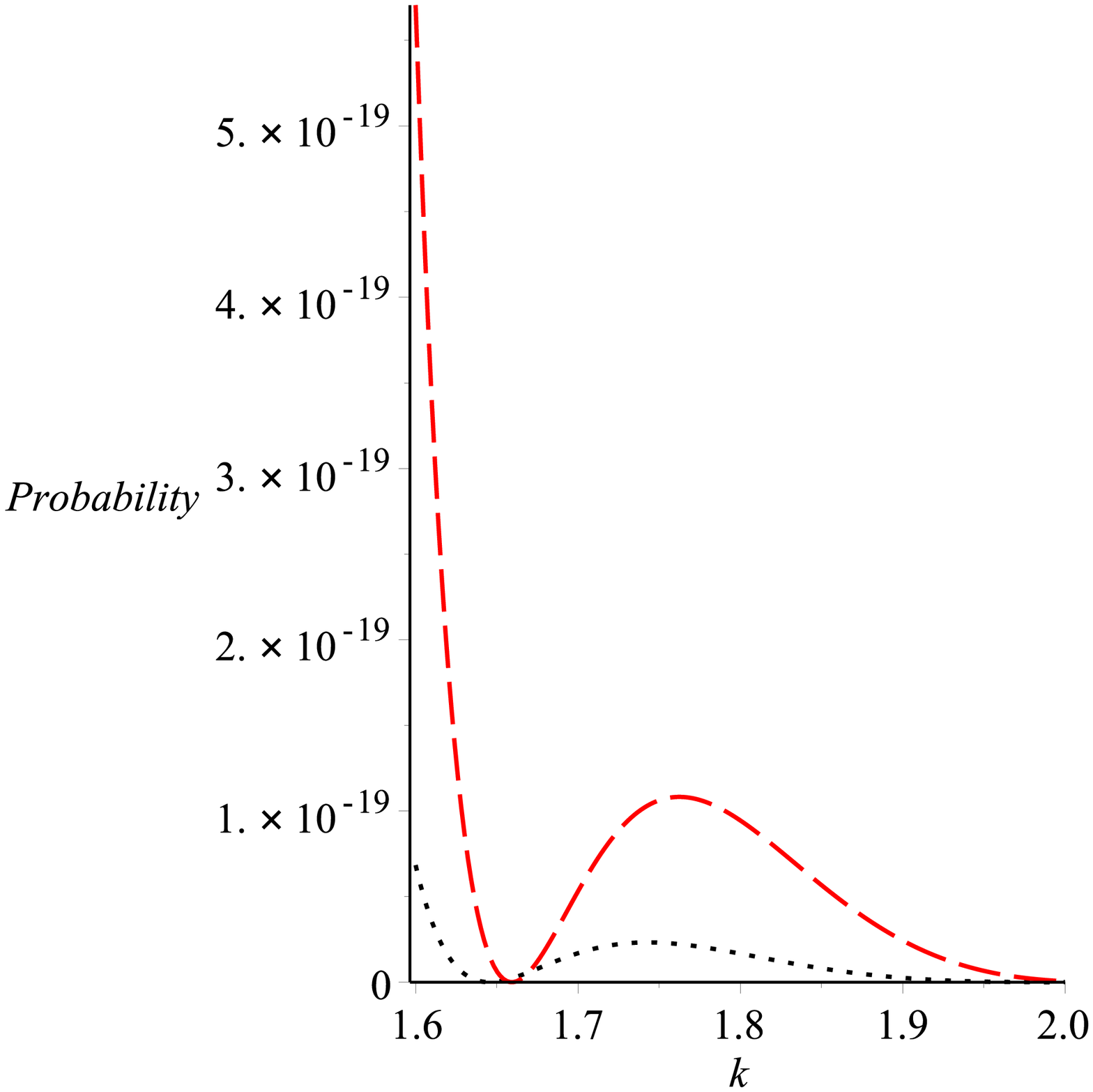}
\caption{$\mathcal{P}$ as a function of $k=\frac{m}{\omega}$. In the left figure the numerical values of the momenta ratios are: $p\,'/p=0.003$  (solid line), $p\,'/p=0.004$ (point line) and $l=2$. In the right figure the numerical values of the momenta ratios are $p\,'/p=0.004$ (point line), $p\,'/p=0.005$ (dashed line) and $l=3$.}
\label{fg4}
\end{figure}

From Figs.(\ref{fg3})-(\ref{fg4}) we observe that, for $p'/p\sim 10^{-3}$ the probability drastically decrease. If we take very small values of the momenta ratios, the probabilities start to have an oscillatory behaviour and tend rapidly to zero.  Figs.(\ref{fg3})-(\ref{fg4}) show that, if we take large values of the quantum number $l$, the probabilities have a pick around $k\sim 1.75$.
Another important observation is that these processes, which involve particle production are significant in the situations in which the expansion factor is large or in other words when the gravitational field is very strong. Therefore, the results obtained in the Cartesian chart $\{t, \vec{x}\}$ regarding the perturbative method for studying the creation of particles leads to the conclusion that, these processes are very probable in the conditions from the early universe, are verified even if we choose to work in another chart of the de Sitter expanding spacetime \cite{1}. Such a result was  obtained by L. Parker many years ago in Refs.\cite{9}\cite{10}, where is introduced the idea that for having nonzero probabilities of particle production is necessary to have a strong gravitational field, corresponding to the conditions of large expansion from early universe. The methods employed by L.Parker in his studies are nonperturbative methods. But another interesting studies regarding the phenomenon of particle production were considered in \cite{020}\cite{0020}. The authors of these papers use perturbative methods and make a comparison between the results of QFT from Minkowski case with those from the de Sitter case. They reach the conclusion that, this effect, relatively to the de Sitter radius is exponentially small. We obtain a similar result in the next section where we show that, if we consider weak gravitational fields the probabilities exponentially decrease. Other papers which discuss the fact that, the particle production phenomenon is important in early universe can be found by consulting Refs.\cite{001}\cite{002}. Our graphs show that, when the Hubble parameter takes small values i.e. $\omega\sim 0$ the probability goes to zero which means that, the Minkowski limit is verified. It is well known that, these processes are forbidden in Minkowski QED because of the simultaneous conservation laws for energy and momentum \cite{024}. But in de Sitter case, although the metric is invariant under space translations we can have situations in which the momentum is no longer a conserved quantity.

Further we want to show the probability dependence as function of the orbital quantum number $l$. The results can be seen in the figures below.
\begin{figure}[H]
\includegraphics[scale=0.35]{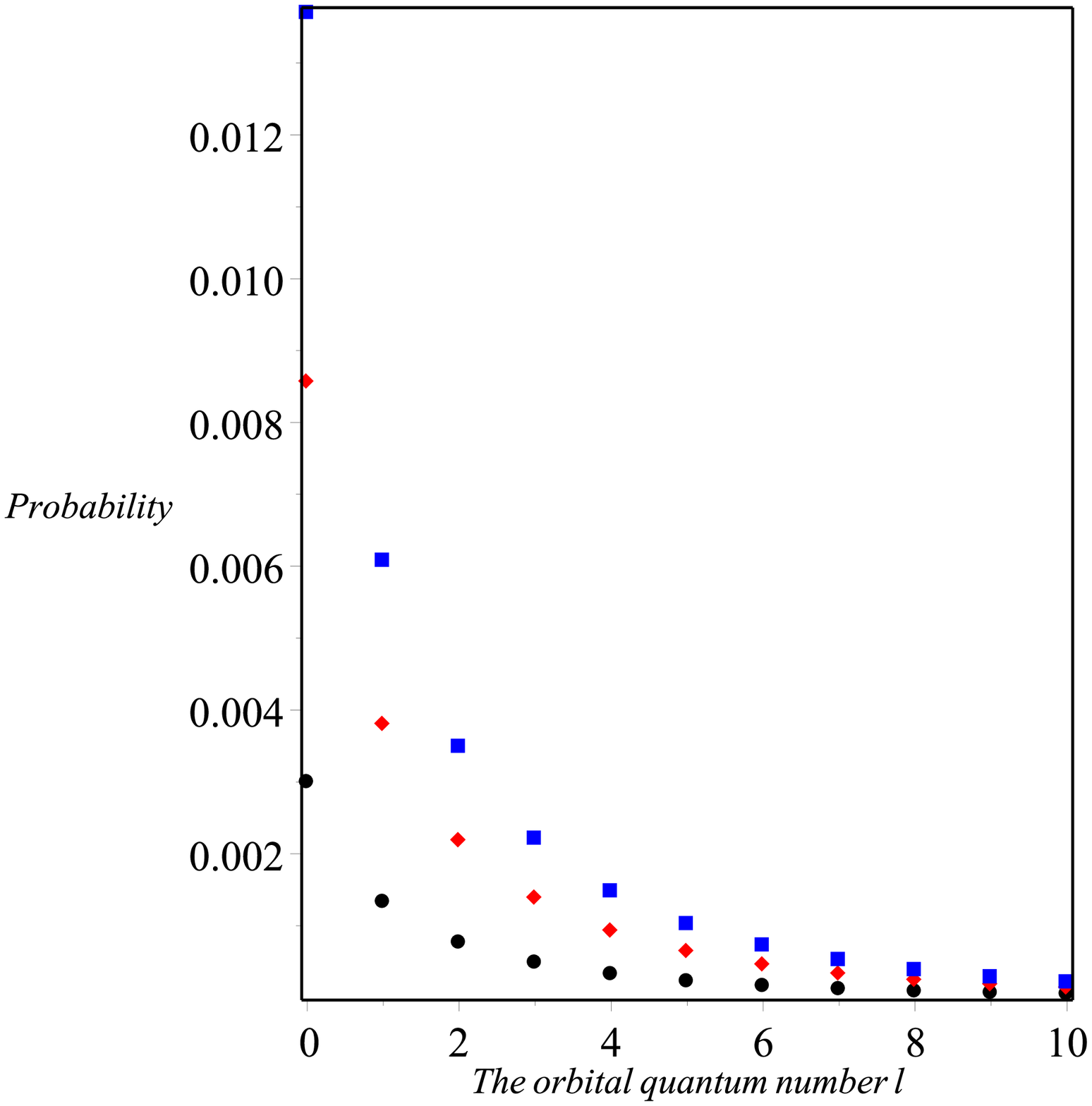}
\includegraphics[scale=0.35]{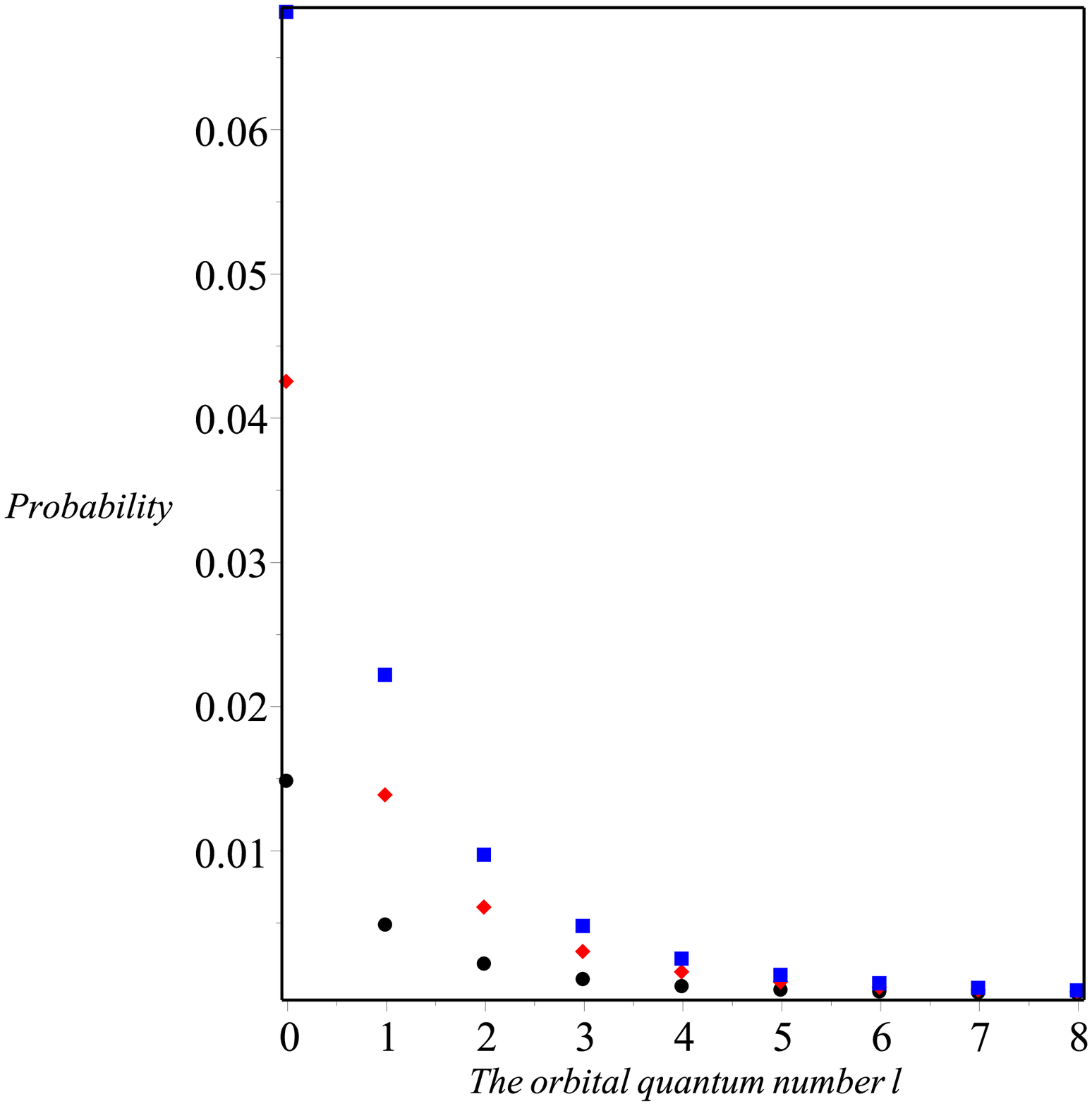}
\caption{$\mathcal{P}$ as a function of the quantum number $l$ for $p\,'/p=0.9$ in the left figure and $p\,'/p=0.8$ in the right figure. Box points are for $k=1.51$, diamond points are for $k=1.6$ and circle points are for $k=1.8$.}
\label{fg5}
\end{figure}

\begin{figure}[H]
\includegraphics[scale=0.35]{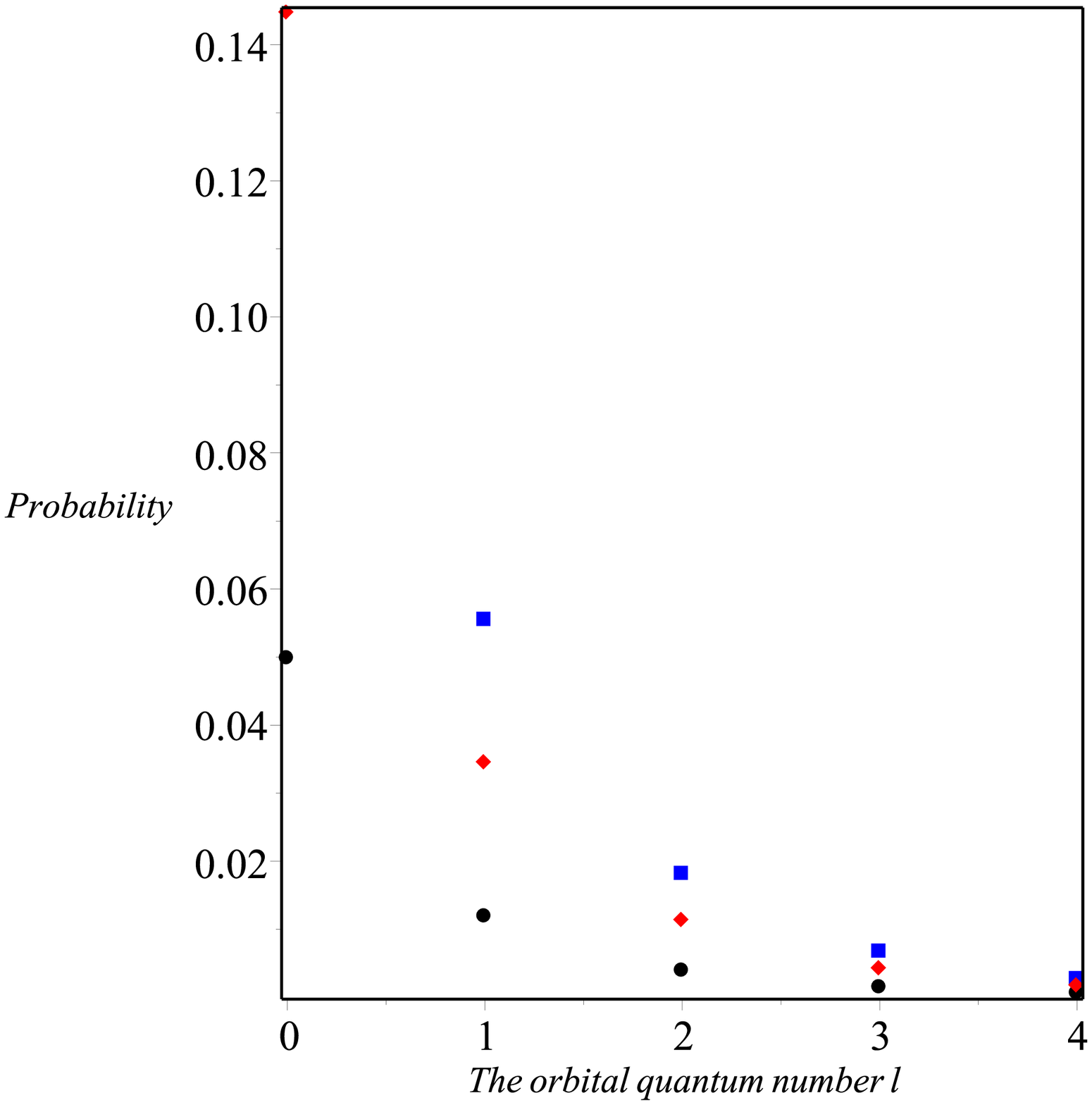}
\includegraphics[scale=0.35]{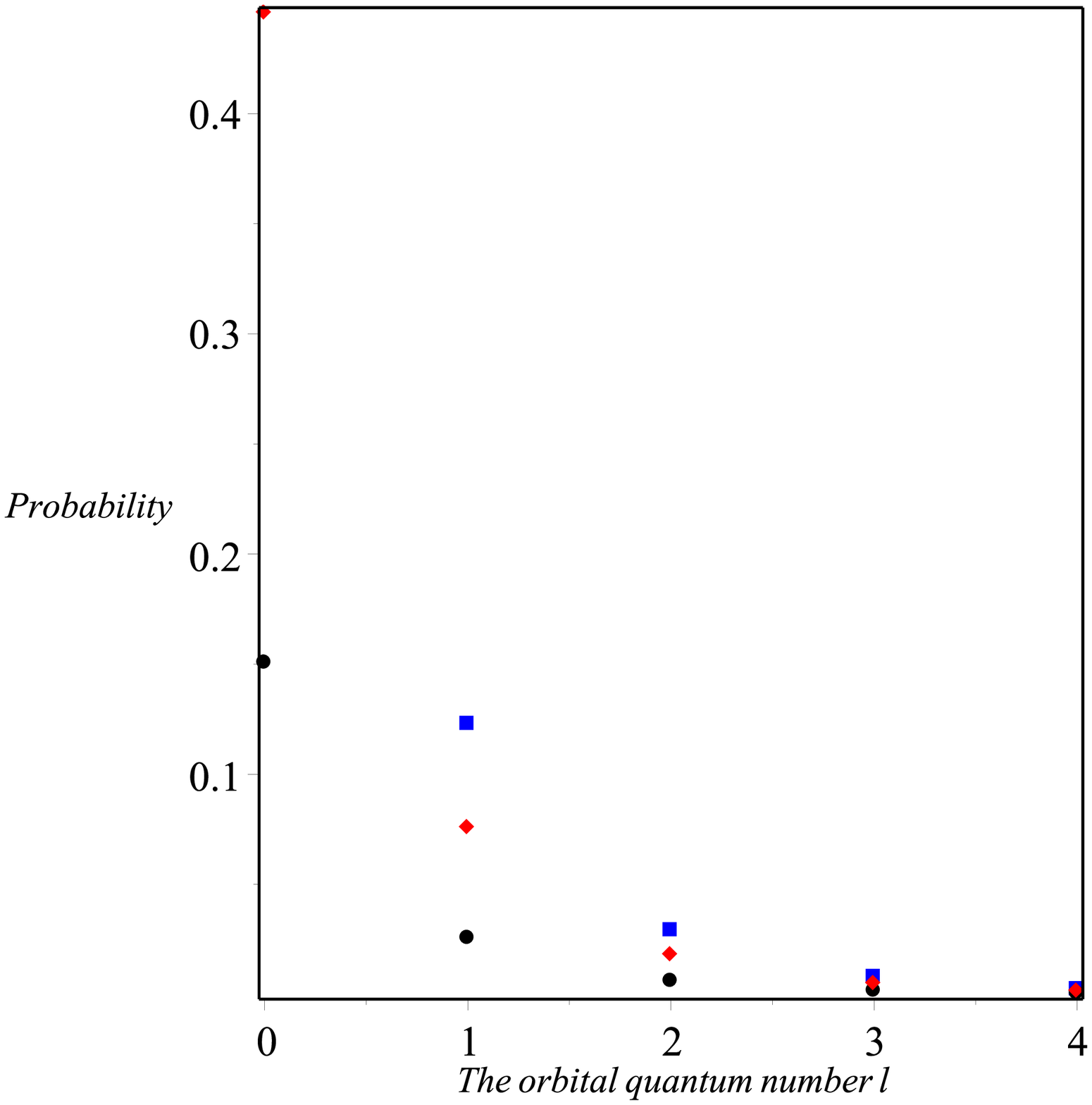}
\caption{$\mathcal{P}$ as a function of the quantum number $l$ for $p\,'/p=0.7$ in left figure and $p\,'/p=0.6$ in the right figure. Box points are for $k=1.51$, diamond points are for $k=1.6$ and circle points are for $k=1.8$.}
\label{fg6}
\end{figure}
Figs.(\ref{fg5})-(\ref{fg6}) show the probability dependence in terms of the orbital quantum number $l$ for fixed values of the parameter $k$ and the momenta ratios $p'/p$. Because the quantum number $l$ can take only discrete values we could evaluate the probability for given values of $l$. From Figs.(\ref{fg5})-(\ref{fg6}) it is more clearly to observe the probability dependence in terms of the quantum number $l$ and it is obvious that, for values $l> (0,\,1)$ the processes of pair production are less probable. It can also be observed that indeed, the most probable transitions are those for which the orbital quantum number is small $(l=0,1)$ and the gravitational field is very strong $m/\omega\sim 3/2$. Another consequence is that, the discrete dependence in terms of $l$ show that, for large values of the quantum number the probability decrease very rapidly, regardless of the gravity field strength. Therefore we can conclude that the dominant processes are those for which the created particles have small angular momentum. The probabilities for large values of $l$, became significant when the momenta of the produced particles is large as well, but even so these probabilities are less important in comparison with the case $l=0$. Now if we have an overview and we look at both sets of graphics we can see the following pattern: $p\,'/p=$ const., $k$ increases, $l$ increases, but the probability decreases.

\section{Weak gravitational field limit}
At this section we want to discuss about the behaviour of the amplitude/probability in the weak gravitational field limit i.e. $\mu=\sqrt{k^2-9/4}\sim k$, for $k=m/\omega\gg 1$.
Namely, we show that, the above conclusions regarding the conditions in which this process has significant probability of pair production, are also verified analytically by the weak gravitational field limit.
Using the relations (\ref{k2}) from Appendix B one can express the functions $f_{k\gg 1}\left(\frac{p'}{p}\right)$, $g_{k\gg 1}\left(\frac{p'}{p}\right)$  that define the amplitude as follows:

\begin{eqnarray}
&&f_{k\gg 1}\left(\frac{p'}{p}\right)=\frac{2\pi}{i\sinh{(\pi k)}}\frac{p^4}{(p^2-p'^2)^2}\left[\left(\frac{p'}{p}\right)^{\frac{1}{2}-ik}+\left(\frac{p'}{p}\right)^{\frac{1}{2}+ik}\right]\nonumber\\
&&+\frac{2\pi}{i\sinh{(\pi k)}}\frac{p^2}{p^2-p'^2}\left[(1-ik)\left(\frac{p'}{p}\right)^{\frac{1}{2}-ik}-(1+ik)\left(\frac{p'}{p}\right)^{\frac{1}{2}+ik}\right]\nonumber\\
\end{eqnarray}
\begin{eqnarray}
g_{k\gg 1}\left(\frac{p'}{p}\right)&=& \frac{2i\pi k}{i\sinh(\pi k)}\frac{p^2}{(p^2-p'^2)}\left[\left(\frac{p'}{p}\right)^{\frac{1}{2}-ik}+\left(\frac{p'}{p}\right)^{\frac{1}{2}+ik}\right]\nonumber\\
&&-\frac{2\pi}{i\sinh(\pi k)}\frac{p^4}{(p^2-p'^2)^2}\left[\left(\frac{p'}{p}\right)^{\frac{5}{2}-ik}-\left(\frac{p'}{p}\right)^{\frac{5}{2}+ik}\right].
\end{eqnarray}
The above relations show that, in the weak gravitational field limit, the amplitude is proportional with $e^{-\pi k}$ and therefore, the probability goes to zero, when $k\rightarrow\infty$ with a term of the form $e^{-2\pi k}$. We mention that, this factor was found in \cite{8}\cite{09}, were is studied the problem of particle production in external fields, in the presence of gravitation, using nonperturbative methods. Therefore, in the Minkowski limit which corresponds to $\omega=0$, the probability of scalar pair production is zero. This result is in accordance with Minkowski QED, where this process is forbidden as a perturbative phenomenon \cite{024}. In the picture discussed here, the Coulomb field, which is a field of weak intensity could lead to particle production only if is coupled with a strong gravitational field. Such conditions could exist in early universe \cite{24}.

\section{Concluding remarks}
In this paper we treated for the first time the problem of charged scalar particles production using the solutions of Klein-Gordon equation with a well defined angular momentum on de Sitter expanding universe. We established the dependence of the probability in terms of the orbital angular momentum.  From these studies is obtained that, the most probable transitions are those which occur at large expansion and the produced particles have the  projection of the angular momentum such that $m= -m'$, which means that the produced particles are spinning  in opposite senses. Since we established the dependence of the probability in terms of the orbital quantum number $l$, we were able to plot the probability as function of discrete  values of $l=0,\,1,\,2,\, ...$ . Our studies show that the most probable transitions occur when the orbital quantum number takes small values. Nevertheless, the plots of the probability as function of the parameter $k=m/\omega$ show that this process is significant in the strong gravitational fields of the early universe. As we mentioned before, even if there are used different methods to study the creation of particles, the various studies for the literature also confirm this conclusion. Another important result obtained here is the fact that, the probability tends to zero for large values of the parameter $k\rightarrow\infty$. This conclusion is verified both by the plots of our functions, but also by the analytical computations of the probability in the weak gravitational field regime. Such a result confirm the predictions from Minkowski QED, which say that, the perturbative method for studying the creation of particles is forbidden by the simultaneous conservation laws for energy and momentum. Also, since the result of the probability depends on the Hubble's constant, it gives us information about the production rate of the present day expansion.

Further we want to address the problem of fermion production with well defined angular momentum, in the presence of an external electromagnetic field. In such a case we can also have information  about the helicity of the created particles and it would be interesting to see which are the cases with the most probable transitions: helicity conservation/helicity nonconservation, depending on the probability dependence of the total angular momentum $\vec{J}= \vec{L}+ \vec{S}$.

\section{Appendix}
In this section we are going to introduce all the technical details related to our studies, namely which are the most important steps in order to compute the amplitude of scalar pair production.

\subsection{ANALYTICAL RESULTS}
\subsubsection{\textbf{The spatial integral}}
The spatial integral depends on products of Bessel J functions and caries information about the angular momentum of the produced particles through the quantum numbers $l, l'$ (see the expression below). The general form of these types of integrals is given in Appendix B, namely the equation $(\ref{j1})$. So after applying (\ref{j1}) our integral, in the case $p>p'$ has the following result:
\begin{eqnarray}\label{s1}
\int_{0}^{\infty}J_{l+\frac{1}{2}}(pr)J_{l'+\frac{1}{2}}(p'r)dr= \frac{1}{p}\,h_{l'}\left(\frac{p'}{p}\right),
\end{eqnarray}
where $h_{l'}\left(\frac{p'}{p}\right)$ represents just a notation and is given by the following expression:
\begin{equation}\label{s2}
h_{l'}\left(\frac{p'}{p}\right)= \left(\frac{p'}{p}\right)^{l'+\frac{1}{2}}\frac{\Gamma(\frac{l+l'+2}{2})}{\Gamma(l'+\frac{3}{2})\Gamma(\frac{l-l'+1}{2})}\,_{2}F_{1}\left(\frac{l+l'+2}{2}, \frac{l'-l+1}{2}; l'+\frac{3}{2}; \frac{p'\,^2}{p^2}\right).
\end{equation}
In order to obtain the result of the spatial integral in the case when $p'>p$ we only have to interchange the momenta of the produced particles and the orbital quantum numbers between them i.e. $p\rightleftarrows p'$; $l\rightleftarrows l'$ in the expressions (\ref{s1})-(\ref{s2}).

\subsubsection{\textbf{The angular integral}}
The angular integral contains information about the angular momentum through the spherical harmonics and can be easily solved using the relation (\ref{u2}) from Appendix B:
\begin{eqnarray}\label{u1}
\int d\Omega \,Y_{l,m}^{*}(\theta,\varphi)Y_{l',m'}^{*}(\theta,\varphi)= (-1)^{m'}\int d\Omega\, Y_{l,m}^{*}(\theta,\varphi)Y_{l',-m'}(\theta,\varphi)=(-1)^{m'}\delta_{l,l'}\delta_{m,-m'}.
\end{eqnarray}
From this result one obtain that the only possible transitions are those for which $l= l'$ and respectively, $m= -m'$.

\subsubsection{\textbf{The temporal integral}}
The result of the transition amplitude (\ref{a3}) show that, the temporal integral depends on Hankel functions, but also depends of the partial derivatives of Hankel functions. This complicated expression is obtained because the amplitude of transition contains the bilateral derivative with respect to time and the time dependence of the quantum modes is contained in the term $e^{-3\omega t/2}$ and in  the arguments of the Hankel functions as well, as one can  see from (\ref{sol2}). When we apply the bilateral derivative operator on (\ref{sol2}), the only remaining terms are those which are proportional to the derivative of Hankel functions. To solve such integrals it is useful to use the recurrence relation (\ref{r1}) between Hankel functions, that is given in Appendix B. Then the temporal integral, in the case $p>p'$ takes the following form:
\begin{eqnarray}\label{t1}
\int_{0}^{\infty} dz z^2 \left[-p'\left[H_{-i\mu}^{(2)}(pz)\cdot H_{-i\mu-1}^{(2)}(p'z)-H_{-i\mu}^{(2)}(pz)\cdot H_{-i\mu+1}^{(2)}(p'z)\right]\right.\nonumber\\
\left.+p\left[H_{-i\mu-1}^{(2)}(pz)\cdot H_{-i\mu}^{(2)}(p'z)-H_{-i\mu+1}^{(2)}(pz)\cdot H_{-i\mu}^{(2)}(p'z)\right]\right].
\end{eqnarray}
The above integrals are solved expressing the Hankel functions in terms of Bessel K functions with the relation (\ref{h1}). Then we arrive at the integrals of the form (\ref{k1}), which can be easily solved in terms of Gamma Euler functions and hypergeometric Gauss functions.
After solving the temporal integrals and the spatial integral as well, we can manipulate the results in order to make the following notations:
\begin{eqnarray}\label{f1}
f_{\mu}\left(\frac{p'}{p}\right)&=&\left(\frac{p'}{p}\right)^{\frac{1}{2}-i\mu}\Gamma(1-i\mu)\Gamma(2+i\mu)\,_{2}F_{1}\left(2,1-i\mu;3; 1-\frac{p'^2}{p^2}\right)\nonumber\\
&& +\left(\frac{p'}{p}\right)^{\frac{1}{2}-i\mu}\Gamma(2-i\mu)\Gamma(1+i\mu)\,_{2}F_{1}\left(1,2-i\mu;3; 1-\frac{p'^2}{p^2}\right).\nonumber\\
\end{eqnarray}
\begin{eqnarray}\label{g1}
g_{\mu}\left(\frac{p'}{p}\right)&=&\left(\frac{p'}{p}\right)^{\frac{1}{2}-i\mu}\Gamma(1-i\mu)\Gamma(2+i\mu)\,_{2}F_{1}\left(1,1-i\mu;3; 1-\frac{p'^2}{p^2}\right)\nonumber\\
&&+\left(\frac{p'}{p}\right)^{\frac{5}{2}-i\mu}\Gamma(2-i\mu)\Gamma(1+i\mu)\,_{2}F_{1}\left(2,2-i\mu;3; 1-\frac{p'^2}{p^2}\right).\nonumber\\
\end{eqnarray}
These functions enter in the definition of the transition amplitude (\ref{a2}) and also in the final expression of the probability (\ref{pr}), with the mention that $f_{\mu}\left(\frac{p}{p'}\right)$ and $g_{\mu}\left(\frac{p}{p'}\right)$ are obtained when we interchange $p\leftrightarrows p'$ in the above expressions.
Using (\ref{hy}) we can express the above functions as follows:
\begin{eqnarray}
f_{\mu}\left(\frac{p'}{p}\right)&=& 2\left[\left(\frac{p'}{p}\right)^{\frac{1}{2}-i\mu}\Gamma(1-i\mu)\Gamma(i\mu)\,_{2}F_{1}\left(2,1-i\mu;1-i\mu; \frac{p'^2}{p^2}\right)\right.\nonumber\\
&&\left.+\left(\frac{p'}{p}\right)^{\frac{1}{2}+i\mu}\Gamma(2+i\mu)\Gamma(-i\mu)\,_{2}F_{1}\left(1,2+i\mu;1+i\mu; \frac{p'^2}{p^2}\right)\right.\nonumber\\
&&\left.+\left(\frac{p'}{p}\right)^{\frac{1}{2}-i\mu}\Gamma(2-i\mu)\Gamma(i\mu)\,_{2}F_{1}\left(1,2-i\mu;1-i\mu; \frac{p'^2}{p^2}\right)\right.\nonumber\\
&&\left.+\left(\frac{p'}{p}\right)^{\frac{1}{2}+i\mu}\Gamma(1+i\mu)\Gamma(-i\mu)\,_{2}F_{1}\left(2,1+i\mu;1+i\mu; \frac{p'^2}{p^2}\right)\right]
\end{eqnarray}

\begin{eqnarray}
g_{\mu}\left(\frac{p'}{p}\right)&=& 2\left[\left(\frac{p'}{p}\right)^{\frac{1}{2}-i\mu}\Gamma(1-i\mu)\Gamma(1+i\mu)\,_{2}F_{1}\left(1,1-i\mu;-i\mu; \frac{p'^2}{p^2}\right)\right.\nonumber\\
&&\left.+\left(\frac{p'}{p}\right)^{\frac{5}{2}+i\mu}\Gamma(2+i\mu)\Gamma(-1-i\mu)\,_{2}F_{1}\left(2,2+i\mu;2+i\mu; \frac{p'^2}{p^2}\right)\right.\nonumber\\
&&\left.+\left(\frac{p'}{p}\right)^{\frac{5}{2}-i\mu}\Gamma(2-i\mu)\Gamma(-1+i\mu)\,_{2}F_{1}\left(2,2-i\mu;2-i\mu; \frac{p'^2}{p^2}\right)\right.\nonumber\\
&&\left.+\left(\frac{p'}{p}\right)^{\frac{1}{2}+i\mu}\Gamma(1+i\mu)\Gamma(1-i\mu)\,_{2}F_{1}\left(1,1+i\mu;i\mu; \frac{p'^2}{p^2}\right)\right].
\end{eqnarray}

\subsection{Useful integrals and formulas}
The spatial integral which contains products of two Bessel functions can be solved with \cite{012}\cite{15}\cite{16}:
\begin{eqnarray} \label{j1}
\int_{0}^{\infty}J_{\mu}(ax)J_{\nu}(bx)dx&=& b^{\nu}a^{-\nu-1}\frac{\Gamma(\frac{\mu+\nu+1}{2})}{\Gamma(\nu+1)\Gamma(\frac{\mu-\nu+1}{2})}\,_{2}F_{1}\left(\frac{\mu+\nu+1}{2},\frac{\nu-\mu+1}{2};\nu+1; \frac{b^2}{a^2}\right)\nonumber\\
&& [a>0, b>0, Re(\mu+\nu)>-1, b<a].
\end{eqnarray}
For getting the final result of the angular integral we use the following relations between the spherical harmonics \cite{012}\cite{15}\cite{16}:
\begin{eqnarray}\label{u2}
Y_{l,m}^{*}(\theta, \varphi)= (-1)^{m} Y_{l,-m}(\theta, \varphi).\nonumber\\
\int d\Omega\, Y_{l,m}^{*}(\theta, \varphi)Y_{l',m'}(\theta, \varphi)= \delta_{l,l'}\delta_{m,m'}.
\end{eqnarray}
The Hankel functions satisfy the following recurrence formula \cite{012}\cite{15}\cite{16}:
\begin{equation}\label{r1}
\frac{dH_{-i\mu}^{(2)}(pz)}{d(pz)}= H_{-i\mu-1}^{(2)}(pz)-H_{-i\mu+1}^{(2)}(pz).
\end{equation}
The Hankel functions can be expressed in terms of Bessel K functions as follows \cite{012}\cite{15}\cite{16}:
\begin{equation}\label{h1}
H^{(2)}_{\nu}(pz)= \left(\frac{2i}{\pi}\right)e^{
i\pi\nu/2}K_{\nu}( ipz).
\end{equation}
The general form of the temporal integral is \cite{012}\cite{15}\cite{16}:
\begin{eqnarray}\label{k1}
&&\int_0^{\infty} dz
z^{-\lambda}K_{\mu}(az)K_{\nu}(bz)=\frac{2^{-2-\lambda}a^{-\nu+\lambda-1}b\,^{\nu}}{\Gamma(1-\lambda)}\Gamma\left(\frac{1-\lambda+\mu+\nu}{2}\right)\Gamma\left(\frac{1-\lambda-\mu+\nu}{2}\right)\nonumber\\
&&\times\Gamma\left(\frac{1-\lambda+\mu-\nu}{2}\right)\Gamma\left(\frac{1-\lambda-\mu-\nu}{2}\right)
\,_{2}F_{1}\left(\frac{1-\lambda+\mu+\nu}{2},\frac{1-\lambda-\mu+\nu}{2};1-\lambda;1-\frac{b^2}{a^2}\right),\nonumber\\
&&Re(a+b)>0\,,Re(\lambda)<1-|Re(\mu)|-|Re(\nu)|.
\end{eqnarray}
The relation between the hypergeometric functions reads \cite{012}\cite{15}\cite{16}:
\begin{eqnarray}\label{hy}
_{2}F_{1}(a,b;c;z)&=& \frac{\Gamma(c)\Gamma(c-a-b)}{\Gamma(c-a)\Gamma(c-b)}\,_{2}F_{1}(a,b;a+b-c+1;1-z)\\ \nonumber
&&+(1-z)^{c-a-b}\,\frac{\Gamma(c)\Gamma(a+b-c)}{\Gamma(a)\Gamma(b)}\,_{2}F_{1}(c-a,c-b;c-a-b+1;1-z).
\end{eqnarray}
In order to obtain the behaviour of the amplitude/probability for values of $k=m/\omega\gg 1$ we use the following formulas \cite{012}\cite{15}\cite{16}:
\begin{eqnarray}\label{k2}
\Gamma(1+z)&=&z\Gamma(z)\nonumber\\
\Gamma(z)\Gamma(-z)&=& -\frac{\pi}{z\sin(\pi z)}\nonumber\\
\Gamma(1-z)\Gamma(z)&=&\frac{\pi}{\sin(\pi z)}\nonumber\\
_{2}F_{1}(a,b,b,z)&=& (1-z)^{-a}.
\end{eqnarray}

\end{document}